# System level analysis of heterogeneous networks under imperfect traffic hotspot localization

Aymen Jaziri, Ridha Nasri and Tijani Chahed

*Abstract*—We study, in this paper, the impact of imperfect small cell positioning with respect to traffic hotspots in cellular networks. In order to derive the throughput distribution in macro and small cells, we firstly perform static level analysis of the system considering a non-uniform distribution of user locations. We secondly introduce the dynamics of the system, characterized by random arrivals and departures of users after a finite service duration, with the service rates and distribution of radio conditions outfitted from the first part of the work. When dealing with the dynamics of the system, macro and small cells are modeled by multi-class processor sharing queues. Macro and small cells are assumed to be operating in the same bandwidth. Consequently, they are coupled due to the mutual interferences generated by each cell to the other. We derive several performance metrics such as the mean flow throughput and the gain, if any, generated from deploying small cells to manage traffic hotspots. Our results show that in case the hotspot is near the macro BS (Base Station), even a perfect positioning of the small cell will not yield improved performance due to the high interference experienced at macro and small cell users. However, in case the hotspot is located far enough from the macro BS, performing errors in small cell positioning is tolerated (since related results show positive gains) and it is still beneficial in offloading traffic from the congested macrocell. The best performance metrics depend also on several other important factors such as the users' arrival intensity, the capacity of the cell and the size of the traffic hotspot.

## I. INTRODUCTION

### A. Motivation and background

WITH the exponential growth of data traffic in modern mobile networks and the emergence of variety of connected devices, the presence of traffic hotspots, reflecting the occurrence of mass events or the existence of areas of capacity bottlenecks, has become one of important scenarios to take into consideration in operators roadmaps to reach their objectives in terms of quality of service (QoS). Therefore, analysis of small cell deployment under the presence of traffic hotspots (i.e. non homogeneous spatial traffic distribution) is a relevant issue to investigate and also to include in network planning process, as it allows to evaluate the efficiency of HetNet[1] deployments

A. Jaziri and R. Nasri are with Orange Labs 38/40 avenue General Leclerc, 92794 Issy-les-Moulineaux, France; (email: aymen.jaziri@orange.com, ridha.nasri@orange.com).

T. Chahed is with Telecom SudParis, 9 rue Charles Fourier, 91011 Evry, France; (email: tijani.chahed@telecom-sudparis.eu).

[1]HetNet = Heterogeneous Network, it is composed of small cells within macrocell coverage. HetNets have been shown to be efficient in improving network performance mostly by covering traffic hotspots and coverage holes.

[1]. Moreover, considering the impact of imperfect small cell positioning provides important results in terms of hotspot localization accuracy beyond which it is worthless to perform network optimization based on the information provided by the applied traffic localization technique. This study falls in the area of network densification and traffic hotspot localization which are among the dominant themes for wireless evolution into 5G networks [2].

### B. Related works

The efficiency of small cell deployment was the subject matter of many works using different network topology models which yields either optimistic or pessimistic results. An excellent survey on heterogeneous cellular networks is elaborated in reference [1]. In [3], [4], authors considered a network with different tiers to evaluate performance metrics such as the coverage probability, the average achievable rate and the average load per tier. The network structure in each tier is based on a spatial Poisson Point Process (PPP). Using PPP model allows for simple and tractable analytical expressions but it remains quite different from the real network layout, especially for macro tier which is close to the hexagonal grid with imperfections [5], [6]. Moreover, the deployment of small cells should be made based on the traffic distribution as well. Considering small cell distribution as a PPP model implies that the traffic hotspot - in case it is the reason for deploying small cells - follows also a PPP model with a constant traffic value in each hotspot which is not the case, as shown in [7]. A different approach was used in [8] where the authors used a fluid model in order to study the impact of small cell location on the performance and Quality of Service (QoS) in HetNets. Non-uniform spatial traffic distribution and user dynamics were however not considered in the analysis.

Performance of cellular networks incorporating sophisticated queuing models have been well investigated in [9]–[13]. In the context of LTE-A HetNets, Ge et al. evaluated the efficiency of deploying femtocells with partially open channels in [12]. The blocking probability and the spectrum and energy efficiency are derived based on a Markov chain model including important parameters used in planning and network design processes such as the number of femtocells, the average number of users and the number of open channels. In reference [13], Saker et al. studied a network of Processor Sharing (PS) queues to model the evolution of the system and how flows are served by either macrocells or small cells. Under the stability condition, capacity, energy consumption and outage rate in the network were evaluated. Authors showed that the densification of the network with small cells is globally a good solution to offload traffic and reduce the energy consumption. However, the spatial traffic distribution was assumed uniform all over the covered area and the small cell deployment did not take into account the presence of traffic hotspots and could be even deployed near the macro BS (Base Station), contrary to what had been recommended in [8]. In fact, it was shown in [8] that deploying small cells in already well-covered areas does not generate capacity gains and even degrades the performance of the network either in the presence of traffic hotspots and also with a uniform spatial traffic distribution (this observation will be also verified in the numerical results of this paper).

*C. Contributions*

In this paper, we firstly derive the throughput distribution following a static level analysis, which refers in our context to a set of decorrelated averaged snapshots of the network where the user location distribution varies only in space. Three different scenarios are studied: i. with macrocells only, ii. with a small cell deployed exactly at the center of the traffic hotspot and iii. with a small cell not perfectly deployed near the traffic hotspot (its position is different from the hotspot one).

We secondly consider the dynamics of the system, with users arriving to the system at random times and leaving it after a finite duration corresponding to the end of their services, with the service rates derived in the first static part of the work. The model in this part is based on coupled multi-class PS queues. The coupling between the macrocell and the small cell is the consequence of the interference generated by each one on the other. We derive analytical expressions for several performance metrics depending on the level of accuracy of traffic hotspot localization tools [14].

From a practical point of view, this study allows to evaluate the gain that can be generated from the deployment of small cells for its own sake and more specifically the performance of realistic and imperfect small cell positioning, due to errors generated by traffic hotspot localization tools. It allows also to identify the tolerated margins of errors in hotspot localization techniques exploited in operational tasks of HetNet design.

*D. Paper structure*

The remainder of this paper is organized as follows. In the next section, we describe the downlink system model with a special focus on radio aspects. In section III, static level analysis is performed considering the three aforementioned scenarios of small cell deployment. Section IV deals with the dynamics of the system, widely named flow level analysis, using results outfitted from section III. Numerical results are highlighted in section V. Section VI concludes the paper and opens directions for future works.

## II. DOWNLINK SYSTEM MODEL

*A. System description*

We consider a cellular network with an infinite number of macrocells, each one transmitting with power level $P$. The location of the macrocells is drawn following a hexagonal grid layout (see Fig. 1) with inter-site distance denoted by $\delta$. We also add a small cell located in $(R_s, \theta_s)$, as illustrated in Fig. 1. The transmit power level of the small cell is $P_s = \kappa P$ with $0 \leq \kappa \leq 1$. A given User Equipment (UE) $m$, with polar coordinates $(r, \theta)$, is served either by the central macrocell (cell at the origin) or the deployed small cell depending on the relative signal strength coming from both antennas. The rest of the cells play the role of interfering cells.

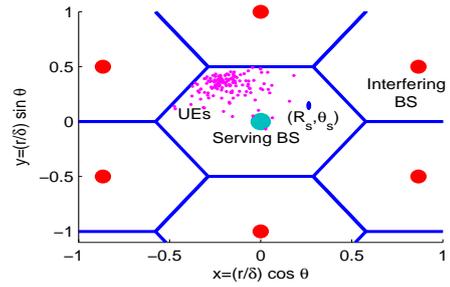

Fig. 1: A snapshot of the network with a traffic hotspot.

In order to evaluate the efficiency of the deployed small cells, we consider a traffic hotspot with polar coordinates $(R_h, \theta_h)$. Without loss of generality, we assume that this hotspot is located inside the central macrocell of the network. This means that $R_h$ is smaller than the radius of the cell $R$. The choice of $R$ is problematic because the disk circumscribed around the hexagon over-estimates interferences whereas the disk inscribed in the hexagon under-estimates interferences and generates coverage holes. For all the theoretical study, we take an arbitrary $R$, such that $R < \delta$, but for the numerical results, we use $R = \delta\sqrt{\frac{\sqrt{3}}{2\pi}}$ which is the radius of the disk having the same area of the hexagon representing the cell. UE locations are spatially distributed in order to form a hotspot centered in $(R_h, \theta_h)$; its measure is given by

$$dt(r,\theta) = \frac{1}{2\pi A^2} e^{-\frac{|re^{i\theta} - R_h e^{i\theta_h}|^2}{2A^2}} r dr d\theta \qquad (1)$$

where $A$ is the standard deviation of the distribution allowing to control the spatial dispersion of the traffic hotspot. An example of the relative spatial distribution is plotted in Fig. 1 with $A = 0.2$ (the same value of $A$ is used in simulations).

**Remark:** Considering more heterogeneity in the spatial traffic distribution is of great importance into the performance evaluation of heterogeneous networks. This aspect and its relative consequences on performance results are discussed in subsection IV-C2. Nevertheless, if we consider more than one traffic hotspot, the impact of imperfect small cell positioning may not make sense since a large error in small cell positioning could help to cover another traffic hotspot. In this way, we may not identify the real impact of imperfect small cell positioning which is the key driver of this work allowing thus to obtain the accuracy level that traffic localization tools must attain. Moreover, with many traffic hotspots, it is not possible to know if it is beneficial to deploy small cells near or far from the macrocell position unless all hotspots are generated in the same region (in the edge or in the center).

As indicated in the introduction, we consider three scenarios: In Scenario 1, the small cell is placed far enough from the traffic hotspot and from the macrocell (i.e. $R_s = \infty$) or it is switched in a sleeping mode (i.e. $P_s = 0$). This scenario represents a benchmark allowing the comparison of a network where small cells contribute to offloading traffic to a network without small cells. Scenario 2 considers a small cell inside the central macrocell that is perfectly deployed in the center of the traffic hotspot, i.e. $(R_h, \theta_h) = (R_s, \theta_s)$. In Scenario 3, we consider a small cell deployed inside the central macrocell as in Scenario 2 but we add imperfections due to the accuracy level of the used traffic hotspot localization method, i.e. $(R_h, \theta_h) \neq (R_s, \theta_s)$.

### B. Channel model and UE association

To model the wireless channel, we consider a distance based pathloss metric with a standard function given by $a|m-C|^{-2b}$, where $|m - C|$ is the distance between the UE $m$ and any cell $C$ in the network, which can be either a macrocell or a small cell, $a$ is a pathloss constant which depends on the type of the environment relative to the type of the cell (indoor, outdoor, rural, urban, etc.) and $2b > 2$ is the pathloss exponent coefficient. Without loss of generality, we consider that the transmit power levels $P$ and $P_s$ include as well the pathloss constant $a$, antenna gain, UE antenna gain and body loss.

Based on the proposed pathloss model, the UE located in $m = (r, \theta)$ is served by the small cell if the RSRP (Reference Signal Received Power) of the small cell is higher than the RSRP coming from the macrocell. This can be expressed by the following inequality

$$P_s |re^{i\theta} - R_s e^{i\theta_s}|^{-2b} > P r^{-2b} \qquad (2)$$

If the constraint in (2) is not satisfied, the UE will be connected to the macrocell.

### C. Link level capacity: Throughput vs SINR

The SINR and the throughput of a UE $m = (r, \theta)$ are respectively denoted by $\gamma(r, \theta)$ and $\eta(r, \theta)$ if received from the macrocell and by $\tilde{\gamma}(r, \theta)$ and $\tilde{\eta}(r, \theta)$ if received from the small cell. Independently from the type of the cell, the relationship between $\gamma$ (in linear scale) and $\eta$ (in Mbps) depends on the UE capability and receiver characteristics, the available bandwidth, the radio conditions and small scale fading effects, the type of the service and the choice of the Modulation and Coding Schemes (MCS). This relationship gives the peak data rate practically experienced by the UE in any position of the network in the absence of any other active user in the cell. It is often modeled by the modified Shannon formula,

$$\eta = \min(K_1 \times W \times \ln(1 + K_2 \times \gamma), \eta_0) \qquad (3)$$

where $K_1$ and $K_2$ are two variables depending on the transmission conditions and can be adapted for each UE speed and category, $W$ is the used bandwidth expressed in MHz and $\eta_0$ is the maximum achievable throughput. For the numerical results, we consider UE category 3 operating at $20 MHz$. From a simple fitting of practical data, we find out that, under laboratory measurement conditions, $\eta_0 = 98 Mbps$, $K_1 = 0.85$ and $K_2 = 1.9$. Fig. 2 provides a comparison between the theoretical and empirical Throughput-SINR link curve of the specified category of UE.

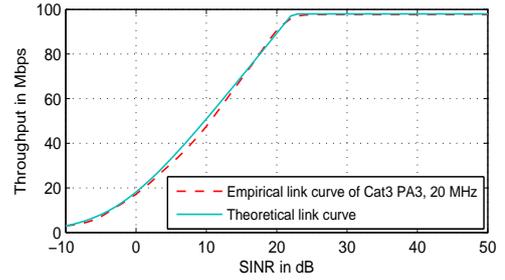

Fig. 2: Throughput-SINR link level curve.

## III. STATIC LEVEL ANALYSIS

In this section, we evaluate the user throughput obtained in each position of the covered region of the studied macrocell and small cell based on the modified Shannon formula given in equation (3). Then, the throughput CCDF (Complementary Cumulative Distribution Function) is calculated considering a user location variable having a measure $dt(r, \theta)$ as given in (1), reflecting the presence of a traffic hotspot in the cell.

**Definition** 1: As mentioned in subsection II-C, $\gamma$ and $\tilde{\gamma}$ are the SINRs received at a UE with polar coordinates $(r, \theta)$ and served respectively by the macrocell and the small cell. Hence, $\gamma$ and $\tilde{\gamma}$ are expressed as follows

$$\gamma(r, \theta) = \frac{1}{g(r) + \kappa |re^{i\theta} - R_s e^{i\theta_s}|^{-2b} r^{2b}}$$

$$\tilde{\gamma}(r, \theta) = \frac{P_s |re^{i\theta} - R_s e^{i\theta_s}|^{-2b}}{(g(r) + 1) P r^{-2b}}$$

where $g(r)$ represents the interference plus noise factor in a network composed of only macrocells. It is defined by the division of the power coming from all the interfering macrocells plus the noise power by the received power from the serving macrocell.

In order to evaluate the impact of an infinite number of interfering macrocells, we established and validated in [6] an expression of $g(r)$ for the considered hexagonal network model. This formula is expressed as follows:

$$g(r) = 6\alpha \left(\frac{r}{\delta}\right)^{2b} \left( \frac{1 + (1-b)^2 \left(\frac{r}{\delta}\right)^2}{\left(1 - \left(\frac{r}{\delta}\right)^2\right)^{2b-1}} + \omega(b) - 1 \right) + \frac{P_N}{P} r^{2b} \qquad (4)$$

where $\alpha$ is the average cell load over all the interfering macrocells, $P_N$ is the noise power level (exterior to the system and including thermal noise) and

$$\omega(b) = 3^{-b} \zeta(b) \left( \zeta(b, \frac{1}{3}) - \zeta(b, \frac{2}{3}) \right)$$

$\zeta(.)$ and $\zeta(., .)$ are respectively the Riemann Zeta and Hurwitz Riemann Zeta functions [15, pp. 1036].

The function $g$ realizes a continuous increasing function from $[0, R]$ to $[0, g(R)]$. And so, It can be explicitly inversed using series reversion or by switching numerically the axes $x$ and $y$. A simple inverse function of $g$ is provided in [16].

**Definition** 2: In the presence of a hotspot (following a spatial distribution such as the one given in (1)) in the region covered by the central macrocell and the deployed small cell $(R_s, \theta_s)$, we define the throughput CCDF for both cells as follows:

$$\mathbb{P}(\eta \geq l) = \frac{1}{S} \int_{S^*} \mathbb{1}\left(P_s |re^{i\theta} - R_s e^{i\theta_s}|^{-2b} \leq Pr^{-2b}\right) \times$$
$$\mathbb{1}\left(\min(K_1 W \ln(1 + K_2 \times \gamma(r,\theta)), \eta_0) \geq l\right) dt(r,\theta) \quad (5)$$

$$\mathbb{P}(\tilde{\eta} \geq l) = \frac{1}{\tilde{S}} \int_{S^*} \mathbb{1}\left(P_s |re^{i\theta} - R_s e^{i\theta_s}|^{-2b} > Pr^{-2b}\right) \times$$
$$\mathbb{1}\left(\min(K_1 W \ln(1 + K_2 \times \tilde{\gamma}(r,\theta)), \eta_0) \geq l\right) dt(r,\theta) \quad (6)$$

where $\mathbb{1}(.)$ is the indicator function which returns 1 if the input condition is verified and zero otherwise and $S^*$ is the studied covered area and

$$S = \int_{S^*} \mathbb{1}\left(P_s |re^{i\theta} - R_s e^{i\theta_s}|^{-2b} \leq Pr^{-2b}\right) dt(r,\theta)$$
$$\tilde{S} = \int_{S^*} \mathbb{1}\left(P_s |re^{i\theta} - R_s e^{i\theta_s}|^{-2b} > Pr^{-2b}\right) dt(r,\theta)$$

with $dt(r,\theta)$ represents the spatial UE location distribution reflecting the presence of a traffic hotspot located in $(R_h, \theta_h)$.

*Lemma 1:* For $l > \eta_0$, the throughput CCDFs in the macrocell and in the small cell are equal to zero since the peak throughput a UE (of the given category) can reach in the best radio conditions, cannot be higher than $\eta_0$.

$$\forall l > \eta_0, \quad \mathbb{P}(\eta \geq l) = 0 \text{ and } \mathbb{P}(\tilde{\eta} \geq l) = 0$$

On the other hand, when $l \leq \eta_0$, the throughput CCDFs are further simplified and are given by

$$\mathbb{P}(\eta \geq l) = \frac{1}{S} \int_{S^*} \mathbb{1}\left(P_s |re^{i\theta} - R_s e^{i\theta_s}|^{-2b} \leq Pr^{-2b}\right)$$
$$\mathbb{1}\left(\frac{1}{\gamma(r,\theta)} \leq \psi(l)\right) dt(r,\theta) \quad (7)$$

$$\mathbb{P}(\tilde{\eta} \geq l) = \frac{1}{\tilde{S}} \int_{S^*} \mathbb{1}\left(P_s |re^{i\theta} - R_s e^{i\theta_s}|^{-2b} > Pr^{-2b}\right)$$
$$\mathbb{1}\left(\frac{1}{\tilde{\gamma}(r,\theta)} \leq \psi(l)\right) dt(r,\theta) \quad (8)$$

with

$$\psi(l) = K_2 \left(e^{\frac{l}{K_1 W}} - 1\right)^{-1} \quad (9)$$

The proof is based on the fact that, the inequality $\min(K_1 W \ln(1 + K_2 \times \tilde{\gamma}(r,\theta)), \eta_0) \geq l$ simplifies to $K_1 W \ln(1 + K_2 \times \tilde{\gamma}(r,\theta)) \geq l$ since $\eta_0 \geq l$. Then, we proceed by performing elementary algebraic transformations.

In the rest of the analysis, we only calculate the throughput CCDF for $l \leq \eta_0$ because it is simply equal to zero otherwise.

### A. Case of no small cells deployed in the studied area

When the macrocell area does not contain small cells, all the UEs in the hotspot, following $dt(r,\theta)$ defined in (1), are served by the macrocell. In this case, the throughput CCDF is given by the following proposition.

*Proposition 1:* When the studied area does not contain a small cell, the throughput CCDF in the macrocell, in the presence of UE location distribution defined as in (1), is expressed by

$$\mathbb{P}(\eta \geq l) = \frac{1}{S} \frac{e^{-\frac{R_h^2}{2A^2}}}{A^2} \int_0^\Lambda r e^{-\frac{r^2}{2A^2}} I_0\left(\frac{rR_h}{A^2}\right) dr \quad (10)$$

with $\Lambda = \min\left(g^{-1}(\psi(l)), R\right)$ and $I_0(.)$ is the first order of the modified Bessel function of the first kind [15, pp. 911], $\psi$ is defined in (9) and $g$ is given in (4).

The proof of this proposition is given in Appendix A.

### B. Case of small cell deployed in the studied area

When the small cell is deployed in the coverage of the macrocell, the throughput CCDF in the macrocell and the small cell are given in the following proposition.

*Proposition 2:* In the presence of a traffic hotspot with a probability measure $dt(r,\theta)$, the throughput CCDFs in the macrocell and in the small cell (with polar coordinates $(R_s, \theta_s)$) are given by the following expressions

$$\mathbb{P}(\eta \geq l) = \frac{1}{S} \int_{\Lambda_0}^{\Lambda} \int_0^{2\pi} \mathbb{1}\left(\cos(\theta - \theta_s) \leq g_2(r)\right) dt(r,\theta) +$$
$$\frac{1}{S} \int_0^{\Lambda_0} \int_0^{2\pi} \mathbb{1}\left(\cos(\theta - \theta_s) \leq g_1(r)\right) dt(r,\theta) \quad (11)$$

$$\mathbb{P}(\tilde{\eta} \geq l) = \frac{1}{\tilde{S}} \int_{\Lambda_0}^{R} \int_0^{2\pi} \mathbb{1}\left(\cos(\theta - \theta_s) \geq g_3(r)\right) dt(r,\theta) +$$
$$\frac{1}{\tilde{S}} \int_0^{\Lambda_0} \int_0^{2\pi} \mathbb{1}\left(\cos(\theta - \theta_s) > g_1(r)\right) dt(r,\theta) \quad (12)$$

with $\Lambda_0 = \min(g^{-1}(\psi(l)-1), R)$, $\Lambda = \min\left(g^{-1}(\psi(l)), R\right)$, $\psi$ and $g$ are given in (9) and (4) respectively.

$$g_1(r) = \frac{R_s^2 + r^2 - \kappa^{1/b} r^2}{2rR_s}$$

$$g_2(r) = \frac{R_s^2 + r^2 - \kappa^{\frac{1}{b}} r^2 (\psi_l - g(r))^{-1/b}}{2rR_s}$$

$$g_3(r) = \frac{R_s^2 + r^2 - (\psi_l \kappa)^{\frac{1}{b}} r^2 (g(r) + 1)^{-1/b}}{2rR_s}$$

The proof of this proposition is given in Appendix B.

**Remark:** The advantage of the obtained expressions is the simplification of the integral with the reduction of the double use of the indicator function in the entire disk to become a simple one in each subarea. In fact, numerical computations of the original expressions in (5) and (6) are time consuming and their simplifications in (10), (11) and (12) reduce significantly the computational costs. This kind of expressions are often implemented in planning and performance prediction tools as a complementary layer to the Monte Carlo simulations module.

## IV. Dynamic level analysis

In this section, we derive performance metrics based on a dynamic level analysis. First, we define the term flow, regularly used in the rest of the analysis, to refer to a session where a user initiates and finishes his transmission successfully. It is characterized by a starting time, corresponding to the time the user arrives to the system, and the size of the file to be transferred. We focus, in this work, on the case of best effort traffic. Flows may belong to different classes where the notion of class reflects the different clusters of users experiencing approximately the same radio conditions. For each class, the arrival rate and the peak data rate are extracted from the static level analysis such as described in the next subsection.

### A. Inputs extracted from the static level analysis

The distribution of users in each flow class can be extracted from the CQI distribution (Channel Quality Indicator distribution) for a cell already installed and operational. However, in the planning task, this information is not available. The throughput distribution, obtained from the static level analysis in the presence of non-uniform traffic, can be therefore a relevant alternative to predict the distribution of radio conditions into the studied area. Following this approach, we extract from the throughput CCDF curve the distribution of radio conditions involving the spatial traffic distribution as follows:

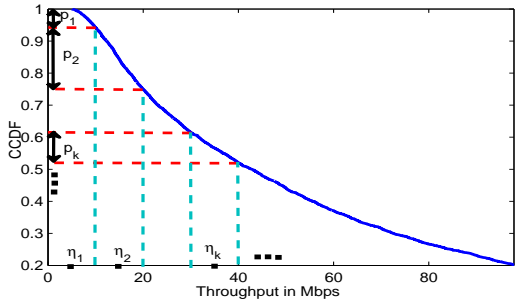

Fig. 3: Extraction of $(p_k, \tilde{p}_l)$ and $(\eta_k, \tilde{\eta}_l)$.

First, we denote by $p_k$ and $\tilde{p}_l$ ($k = 1..K$ and $l = 1..L$) the proportion of users in each class of radio conditions where $K$ and $L$ represent the number of classes in the macrocell and the small cell respectively. Second, in Fig. 3, the x-axis of the throughput is discretized in order to divide users into classes of flows according to their peak data rate. Each class is characterized by $\eta_k$ for the macrocell and $\tilde{\eta}_l$ for the small cell which corresponds to the mean (or the max [9], [11]) of peak data rates in this class. Third, the proportion of users $p_k$ in class $k$ (or $p_l$ in class $l$ for the small cell) is directly extracted from the y-axis by simply taking the value of $\mathbb{P}(\eta_{min,k} \leq \eta \leq \eta_{max,k})$, where $\eta_{min,k}$ and $\eta_{max,k}$ represent the boundaries of the interval of class $k$ peak data rate.

### B. Traffic characteristics and dynamic system model

For both cells, random arrivals of data flows are modeled by multi-class flows; the flow classes are sorted according to the associated peak data rate $\eta_k$ (or $\tilde{\eta}_l$ for small cell) in an increasing order. All users of the same class have approximately the same data rate when scheduled in the same cell, however the data rate differs from a flow class to another one. We assume that each flow class arrival follows a Poisson Process of intensity $\lambda_k = \lambda p_k = \lambda_{Tot} \frac{S}{S+\tilde{S}} p_k$ in the macrocell and $\tilde{\lambda}_l = \tilde{\lambda}\tilde{p}_l = \lambda_{Tot} \frac{\tilde{S}}{S+\tilde{S}} \tilde{p}_l$ in the small cell, where $\lambda_{Tot}$ is the total arrival intensity in the studied system.

Flow sizes, denoted by $\sigma$, are assumed to be mutually independent and exponentially distributed with mean $\sigma_0$ in Mbits. Traffic is assumed to be FTP-like data. Such elastic flows result from the transfer of electronic documents like peer-to-peer file sharing, typically characterized by a file size and a variable duration which depends on the network capacity and load, and also on the radio conditions of the user.

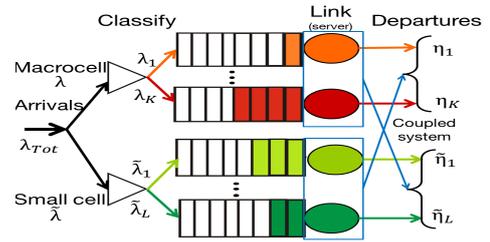

Fig. 4: Dynamic system model.

Cell capacity is assumed to be shared between all transmissions according to a multi-class $M/G/1$ Processor Sharing discipline (with a round robin scheduling policy) wherein resources are equally shared between all the users and all the classes. Moreover, the service in the macrocell and the small cell are coupled because the peak data rate in each class of the macrocell users and small cell ones depends on the existence of an active user in the small cell and in the macrocell, respectively. Indeed, macrocell interference on small cell users is accounted for when there is at least one active user in the macrocell. Similarly, small cell interference on macrocell users is accounted for when there is at least one active user in the small cell. Therefore, $\eta_k$ and $\tilde{\eta}_l$ take each one two values depending on the presence of an active user in the interfering cell. Their expressions are given as follows

$$\forall k = 1..K, \ \eta_k = \begin{cases} \eta_{k,0} \text{ if no user is served by the small cell} \\ \eta_{k,1} \text{ otherwise} \end{cases} \quad (13)$$

$$\forall l = 1..L, \ \tilde{\eta}_l = \begin{cases} \tilde{\eta}_{l,0} \text{ if no user is served by the macrocell} \\ \tilde{\eta}_{l,1} \text{ otherwise} \end{cases} \quad (14)$$

where $\eta_{k,0}$ is given by the throughput distribution curve of subsection III-A (equation (10)) and $\eta_{k,1}$ is given in subsection III-B (equation (11)). For the small cell, $\tilde{\eta}_{l,0}$ is simply given by deleting the contribution of the macrocell in the interference experienced by the small cell's UEs. $\tilde{\eta}_{l,1}$ is also derived from the throughput distribution curve of subsection III-B (equation (12)). The dynamic system model is depicted in Fig. 4.

## C. Performance analysis

### 1) Preliminary analysis of the model

We denote by $\rho$ and $\tilde{\rho}$ the load of the macrocell and the small cell respectively. In the quasi-stationary regime, i.e. the rate processes evolve on an infinitely slow time scale, $\rho$ and $\tilde{\rho}$ are given by

$$\rho = \sum_{k=1}^{K} \lambda_k \mathbb{E}\left(\frac{\sigma}{\eta_k}\right) \ , \ \tilde{\rho} = \sum_{l=1}^{L} \tilde{\lambda}_l \mathbb{E}\left(\frac{\sigma}{\tilde{\eta}_l}\right) \quad (15)$$

At instant $t$, we denote by $n_k(t)$ and $\tilde{n}_l(t)$ the number of flows in each class of the macrocell and the small cell respectively. Then, $(n(t), \tilde{n}(t)) = (n_k, \tilde{n}_l)_{k=1..K, l=1..L}$ represents an irreducible Markov process (Fig. 5) with non-null transition rates from state $(n, \tilde{n})$ to state $(m, \tilde{m})$ given by

$$q((n,\tilde{n}), (n+e_k, \tilde{n})) = \lambda_k$$
$$q((n,\tilde{n}), (n-e_k, \tilde{n})) = \mathbb{1}(n_k > 0)\frac{\eta_k}{|n|}$$
$$q((n,\tilde{n}), (n, \tilde{n}+e_l)) = \tilde{\lambda}_l$$
$$q((n,\tilde{n}), (n, \tilde{n}-e_l)) = \mathbb{1}(\tilde{n}_l > 0)\frac{\tilde{\eta}_l}{|\tilde{n}|} \quad (16)$$

where the states $n = (n_1, .., n_K)$ and $\tilde{n} = (\tilde{n}_1, .., \tilde{n}_L)$ respectively in the macrocell and the small cell represent the number of simultaneous active flows in each flow class. We denote by $|n| = \sum_{k=1}^{K} n_k$ and $|\tilde{n}| = \sum_{l=1}^{L} \tilde{n}_l$ the cardinality of all the active flows in the macrocell and the small cell respectively and $e_k$ or $e_l$ are the row vectors with all components equal to 0 except the $k^{th}$ or $l^{th}$ one equal to 1.

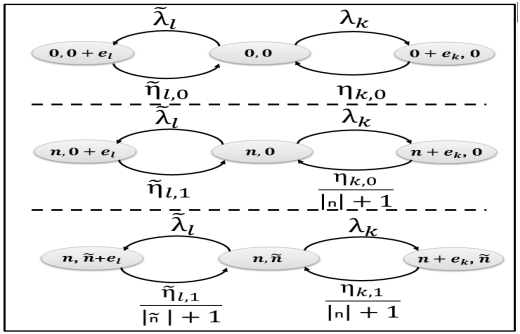

Fig. 5: Markov process of the Macro-Small cells system.

The transition from state $(n, \tilde{n})$ to $(n+e_k, \tilde{n})$ means that a new macrocell user arrived with radio conditions of class $k$. Likewise, the transition from state $(n, \tilde{n})$ to $(n, \tilde{n}+e_l)$ means that a new small cell user arrived with radio conditions of class $l$ characterized by its arrival rate $\tilde{\lambda}_l$ and its service data rate $\tilde{\eta}_l$. However, transition from state $(n, \tilde{n})$ to $(n-e_k, \tilde{n})$ signifies that a user of class $k$ attached to the macrocell had successfully finished his service. The same applies for the transition from $(n, \tilde{n})$ to $(n, \tilde{n}+e_l)$ where a user in class $l$ in the small cell completed normally its transmission and quit the system.

Furthermore, we take

$$q((n,\tilde{n}), (n,\tilde{n})) = - \sum_{(m,\tilde{m}) \neq (n,\tilde{n})} q((n,\tilde{n}), (m,\tilde{m})) \quad (17)$$

Knowing that $\rho, \tilde{\rho} < 1$, there exists a unique invariant probability distribution $\pi_{n,\tilde{n}}, (n, \tilde{n}) \in \mathbb{Z}_+^K \times \mathbb{Z}_+^L$ satisfying

$$\pi Q = 0 \quad and \quad \sum_{(n,\tilde{n}) \in \mathbb{Z}_+^K \times \mathbb{Z}_+^L} \pi_{n,\tilde{n}} = 1 \quad (18)$$

where $\pi = [\pi_{n,\tilde{n}}]_{(n,\tilde{n}) \in \mathbb{Z}_+^K \times \mathbb{Z}_+^L}$ is the row vector composed of the stationary probabilities in each state and $Q = [q((n,\tilde{n}), (m,\tilde{m}))]_{n,m \in \mathbb{Z}_+^K \ and \ \tilde{n},\tilde{m} \in \mathbb{Z}_+^L}$ is the matrix of transition rates.

The relative balance equations of the defined Markov process can be also given by

$$\sum_{k=1}^{K} \left( \mathbb{1}(n_k > 0)\lambda_k \pi_{n-e_k, \tilde{n}} + \frac{\eta_k}{|n|+1}\pi_{n+e_k, \tilde{n}} \right)$$
$$+ \sum_{l=1}^{L} \left( \mathbb{1}(\tilde{n}_l > 0)\tilde{\lambda}_l \pi_{n, \tilde{n}-e_l} + \frac{\tilde{\eta}_l}{|\tilde{n}|+1}\pi_{n, \tilde{n}+e_l} \right) = \pi_{n,\tilde{n}}$$
$$\left( \sum_{k=1}^{K} \mathbb{1}(n_k > 0)\frac{\eta_k}{|n|} + \lambda_k + \sum_{l=1}^{L} \mathbb{1}(\tilde{n}_l > 0)\frac{\tilde{\eta}_l}{|\tilde{n}|} + \tilde{\lambda}_l \right) \quad (19)$$

### 2) Resolution of the model

The interaction between the macrocell and the small cell with a multi-class PS system makes an exact analysis very complex and impractical since the derivation of the exact solution to the system of equations defined in (18) or also in (19) is very difficult. Therefore, we propose, for tractability issues, to transform the system model as described next.

***Lemma 2:*** Under the stability condition, i.e. $\rho, \tilde{\rho} < 1$, the average loads in (15) is given by

$$\rho = 1 - P_0 \ , \ \tilde{\rho} = 1 - \tilde{P}_0 \quad (20)$$

where $P_0$ and $\tilde{P}_0$ are the probabilities of the steady states where no user is in the macrocell and the small cell respectively.

**Proof:** As largely considered in literature, the load is equal to the percentage of time where cell's resources are occupied by UEs. In the studied model, if at least one UE is active, all the available resources in the cell will be occupied by it or shared with other UEs since the traffic generated is elastic. Based on this scheduling policy, the load is equal to the percentage of time where there was at least one active flow in the cell. Hence, $\rho = 1 - P_0, \tilde{\rho} = 1 - \tilde{P}_0$. ∎

In the studied system, macrocell flows experience permanent interference-limited service rate $\eta_{k,1}$ during a proportion $1 - \tilde{P}_0$ of the total observation time and interference-free service rate $\eta_{k,0}$ during the rest of the observation time (a proportion of $\tilde{P}_0$). The same applies for the small cell. It follows that, based on the expression of the load given in (15) and under the stability condition, $\rho$ becomes as follows

$$\rho = \sum_{k=1}^{K} \left[ \tilde{P}_0 \frac{\lambda_k \sigma_0}{\eta_{k,0}} + \left(1 - \tilde{P}_0\right) \frac{\lambda_k \sigma_0}{\eta_{k,1}} \right] \quad (21)$$

Similarly, the load $\tilde{\rho}$ in the small cell is given by

$$\tilde{\rho} = \sum_{l=1}^{L} \left[ P_0 \frac{\tilde{\lambda}_l \sigma_0}{\tilde{\eta}_{l,0}} + (1-P_0) \frac{\tilde{\lambda}_l \sigma_0}{\tilde{\eta}_{l,1}} \right] \quad (22)$$

***Theorem 1:*** In the coupled multi-class processor sharing system defined above, the stationary probability of having at least one active flow in the macrocell (in the small cell respectively) is equal to the cell load. The latter is given by

$$\rho = \frac{\sum_{k=1}^{K} \left[ \frac{\lambda_k \sigma_0}{\eta_{k,0}} + \left( \frac{\lambda_k \sigma_0}{\eta_{k,1}} - \frac{\lambda_k \sigma_0}{\eta_{k,0}} \right) \times \sum_{l=1}^{L} \left[ \frac{\tilde{\lambda}_l \sigma_0}{\tilde{\eta}_{l,0}} \right] \right]}{1 - \sum_{k=1}^{K} \left( \frac{\lambda_k \sigma_0}{\eta_{k,1}} - \frac{\lambda_k \sigma_0}{\eta_{k,0}} \right) \times \sum_{l=1}^{L} \left( \frac{\tilde{\lambda}_l \sigma_0}{\tilde{\eta}_{l,1}} - \frac{\tilde{\lambda}_l \sigma_0}{\tilde{\eta}_{l,0}} \right)} \quad (23)$$

$$\tilde{\rho} = \frac{\sum_{l=1}^{L} \left[ \frac{\tilde{\lambda}_l \sigma_0}{\tilde{\eta}_{l,0}} + \left( \frac{\tilde{\lambda}_l \sigma_0}{\tilde{\eta}_{l,1}} - \frac{\tilde{\lambda}_l \sigma_0}{\tilde{\eta}_{l,0}} \right) \times \sum_{k=1}^{K} \left[ \frac{\lambda_k \sigma_0}{\eta_{k,0}} \right] \right]}{1 - \sum_{l=1}^{L} \left( \frac{\tilde{\lambda}_l \sigma_0}{\tilde{\eta}_{l,1}} - \frac{\tilde{\lambda}_l \sigma_0}{\tilde{\eta}_{l,0}} \right) \times \sum_{k=1}^{K} \left( \frac{\lambda_k \sigma_0}{\eta_{k,1}} - \frac{\lambda_k \sigma_0}{\eta_{k,0}} \right)} \quad (24)$$

with $\sigma_0$, $\lambda_k$, $\tilde{\lambda}_l$, $\eta_{k,0}$, $\tilde{\eta}_{l,0}$, $\eta_{k,1}$ and $\tilde{\eta}_{l,1}$ are defined as in subsection IV-B.

**Proof:** In order to find (23), we take, at first, the expression of the load $\rho$ given in (21). Based on the equality given in Lemma 2, we replace $\tilde{P}_0$ by $1 - \tilde{\rho}$ in (21). Then, $\tilde{\rho}$ is also replaced by its expression given in (22) and we obtain the following identity, equivalent to (23),

$$\rho = \sum_{k=1}^{K} \Big[ \frac{\lambda_k \sigma_0}{\eta_{k,0}} + \left( \frac{\lambda_k \sigma_0}{\eta_{k,1}} - \frac{\lambda_k \sigma_0}{\eta_{k,0}} \right) \times$$
$$\sum_{l=1}^{L} \left[ \frac{\tilde{\lambda}_l \sigma_0}{\tilde{\eta}_{l,0}} + \left( \frac{\tilde{\lambda}_l \sigma_0}{\tilde{\eta}_{l,1}} - \frac{\tilde{\lambda}_l \sigma_0}{\tilde{\eta}_{l,0}} \right) \rho \right] \Big] \quad (25)$$

Transforming equation (25) to extract $\rho$ yields (23). The same steps are followed to obtain (24). ∎

Based on the expressions given in (23) and (24), the stationary distribution is derived in the following proposition.

***Proposition 3:*** Under the stability condition ($\rho, \tilde{\rho} < 1$), the stationary distribution of the state $(n, \tilde{n})$ becomes equal to

$$\pi_{n,\tilde{n}} = \frac{|n|!(1-\rho)}{\prod_{k=1}^{K} (\tilde{\rho} n_k)! \left((1-\tilde{\rho})n_k\right)!} \frac{|\tilde{n}|!(1-\tilde{\rho})}{\prod_{l=1}^{L} (\rho \tilde{n}_l)! \left((1-\rho)\tilde{n}_l\right)!} \times$$
$$\prod_{k=1}^{K} \frac{(\lambda_k \sigma_0)^{n_k}}{\eta_{k,0}^{(1-\tilde{\rho})n_k} \eta_{k,1}^{\tilde{\rho} n_k}} \prod_{l=1}^{L} \frac{(\tilde{\lambda}_l \sigma_0)^{\tilde{n}_l}}{\tilde{\eta}_{l,0}^{(1-\rho)\tilde{n}_l} \tilde{\eta}_{l,1}^{\rho \tilde{n}_l}} \quad (26)$$

**Proof:** For each class $k$ of flows in the macrocell, the queuing system can be modeled with a proportion $\tilde{P}_0$ of flows experiencing a peak data rate equal to $\eta_{k,0}$ and a proportion $1-\tilde{P}_0$ of flows experiencing a peak data rate equal to $\eta_{k,1}$. The same applies for the small cell. The system is partially modified and behaves like a network with two types of classes in both the macrocell and the small cell: macro and small cell flow classes suffering all the time from interference coming from the small cell and the macrocell respectively and flow classes not affected by this interference. Consequently, the service rate in each cell does not depend anymore on the queue state of the other interfering cell. Under the stability condition and following [17, Chap. 7, p. 119], the stationary distributions are directly derived as in (26), which completes the proof. ∎

**Remark:** Considering more heterogeneity in the traffic distribution can be depicted by two main possible scenarios: The first one consists in adding a uniform traffic distribution in addition to the hotspot distribution. In this case, analysis are not significantly altered since a uniform traffic will approximately add the same proportion of users in each class of flows which leads to obtain the same results but for higher value of $\lambda_{Tot} = \lambda_{Hotspot} + \lambda_{Uniform}$. The second scenario consists in considering more than one traffic hotspot (but not too many because otherwise the traffic becomes as if uniformly distributed in the macrocell). Despite its incapability to respond to the main objective of this paper for the reasons cited in subsection II-A, the scenario of more than one traffic hotspot can be a particular case of our study and simplify the system mainly for the dynamic level analysis. For instance, if the deployed small cell is serving a traffic hotspot and the macrocell is serving other traffic hotspots, then the load in the macrocell will tend to 1. Consequently, the coupling defined in (14) will disappear since there will be always at least one active user in the macrocell.

From (26) and following [11], the mean number of flows in each class is equal to

$$N_k = \tilde{\rho} \frac{\lambda_k \sigma_0}{(1-\rho)\eta_{k,1}} + (1-\tilde{\rho}) \frac{\lambda_k \sigma_0}{(1-\rho)\eta_{k,0}} \quad (27)$$

$$\tilde{N}_l = \rho \frac{\tilde{\lambda}_l \sigma_0}{(1-\tilde{\rho})\tilde{\eta}_{l,1}} + (1-\rho) \frac{\tilde{\lambda}_l \sigma_0}{(1-\tilde{\rho})\tilde{\eta}_{l,0}} \quad (28)$$

Applying Little's law to the presented ergodic process provides that the mean service time of class $k$ in the macrocell is $\tau_k = \frac{N_k}{\lambda_k}$, whereas for class $l$ in the small cell is $\tilde{\tau}_l = \frac{\tilde{N}_l}{\tilde{\lambda}_l}$.

Let $\upsilon_k$ and $\tilde{\upsilon}_l$ be the stationary throughput of class $k$ in the macrocell and class $l$ in the small cell, respectively. Then, from (21), (22) and the expression of service time in each class, we have the following identities:

$$\upsilon_k = \tilde{\rho}\eta_{k,1}(1-\rho) + (1-\tilde{\rho})\eta_{k,0}(1-\rho) \quad (29)$$
$$\tilde{\upsilon}_l = \rho\tilde{\eta}_{l,1}(1-\tilde{\rho}) + (1-\rho)\tilde{\eta}_{l,0}(1-\tilde{\rho}) \quad (30)$$

According to the previous expressions, when the load is very small in the macrocell, the flow throughput $\upsilon_k$ is at its maximum. It then decreases linearly with its own cell load. When the load is very small in the small cell, the flow throughput in the macrocell is at its maximum and decreases linearly with the load in the small cell.

## V. NUMERICAL RESULTS

TABLE I: Parameters' configuration.

| | |
|---|---|
| Macro deployment | infinite hexagonal with $\delta = 1$ km |
| Pathloss model MtoUE | $151 + 37.6 \log_{10}(d_{km})$ |
| Pathloss model StoUE | $148 + 36.7 \log_{10}(d_{km})$ |
| BS power | Macro:46dBm, Small:30dBm |
| Antenna gain with cable loss | Macro:18dBi, Small:6dBi |
| Frequency/Bandwidth | 2.6 Ghz / 20 Mhz |
| File size/Scheduling/Traffic type | 2Mb / PS (Round Robin Scheduling) / FTP |

We now evaluate the different performance metrics of the studied scenarios with static and dynamic levels. We show in Table I the most relevant parameters used in the numerical evaluation (taken from the 3GPP LTE-Advanced specifications [18] or also from operational settings).

In the first simulation, we vary the position of the hotspot by varying $R_h$ and determine the minimum distance to the macro BS from which the small cell deployment can be considered as an efficient solution. We assume the error of small cell positioning to be constant relative to the position of the hotspot, taking the following values: error of 0 meters (perfect hotspot localization), 60 meters (accuracy provided in [14]) and 120 meters (current accuracy when using probes) between the variables $R_h$ and $R_s$ and we suppose that $\theta_s$ is equal to $\theta_h$ (equal to $\frac{\pi}{3}$). Then, the mean user throughputs is calculated as a function of $R_h$.

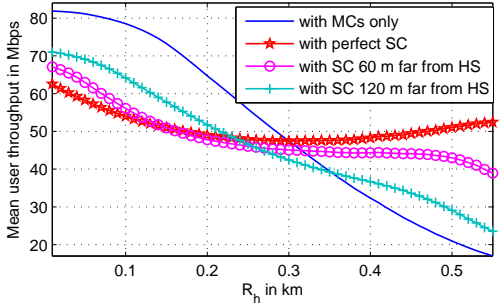

Fig. 6: The mean user throughput for different locations of the traffic hotspot with varying $R_h$.

From Fig. 6, we observe that deploying a small cell near the macro BS does not generate additional capacity gains since the interference in this case is very high in comparison to the SNR received either from the serving small cell or macrocell. In fact, for a hotspot in a position less than 300 meters from the macro BS, the evaluation of the impact of bad localization of the traffic hotspot is worthless and not justified, even with a perfect positioning. Hence, the deployment of a small cell near the macro BS does not help to offload the traffic and deteriorates the throughput in the macrocell. However, the deployment of a small cell improves significantly the overall performance of the macrocell in the presence of the hotspot in cell edge. In the latter case, the small cell still generates positive offloading gains even when its position does not match exactly with the position of the traffic hotspot.

In the rest of the numerical results, we consider a specific scenario ($R_h = 0.5, \theta_h = \frac{\pi}{3}$) where the traffic hotspot is in area where the deployment of a small cell is more eligible to generate positive gains. Like in the previous results, we fix an error of 0 meters, 60 meters and 120 meters respectively between the variables $R_h$ and $R_s$. $\theta_s$ and $\theta_h$ are taken equal to $\frac{\pi}{3}$ since the small cell deployment is practically insensitive to the angle coordinate as we verified in [6].

*A. Static level results*

The throughput CCDF is plotted for the different proposed scenarios in Fig. 7. In legends, MC and SC mean macrocell and small cell respectively, HS means the traffic hotspot and perfect SC means that the small cell is deployed in the center of the hotspot (perfect in terms of positioning).

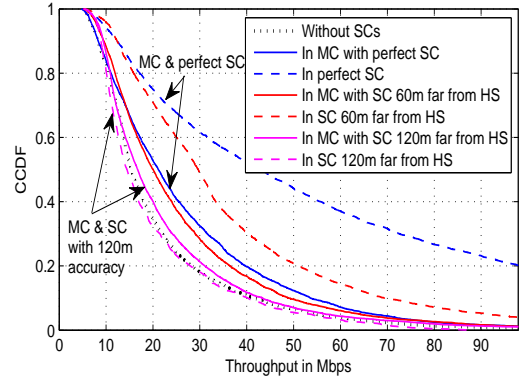

Fig. 7: Throughput CCDF in the 3 scenarios defined in the static level study.

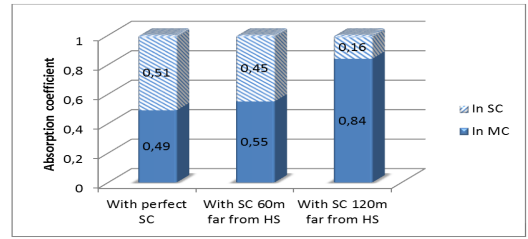

Fig. 8: The absorption coefficient.

From Fig. 7, we observe that the perfect positioning of the deployed small cell improves significantly the capacity of the system (composed of the macrocell and the small cell deployed within its covered area) since user locations with degraded throughputs will be covered by the small cell and hence will experience a better SINR level. Moreover, we note that, even with taking into consideration the imperfect traffic hotspot localization (60 meters), the system capacity is improved comparing to the scenario where only macrocells are operating in the studied area. However, when the distance between the small cell and the hotspot increases due to inaccurate traffic hotspot localization technique, the solution of deploying small cells becomes useless because the cell throughput distribution is not improved anymore compared to a network without small cells. Based on these results, we conclude that the efficiency of small cell deployment mainly depends on the precision of the traffic hotspot localization process in addition to the position of the hotspot with respect to the macrocell's position.

We plot in Fig. 8 the absorption coefficient defined as $S/(S+\tilde{S})$. This coefficient allows to determine the percentage of user locations, with considering the presence of the traffic hotspot, that can be covered and served by the small cell. We observe that this ratio is high when the small cell is perfectly deployed in the center of the traffic hotspot. However, when the distance between the small cell position and the hotspot center increases, less traffic is carried by the small cell.

## B. Dynamic level results

Based on the results obtained from the static level analysis, we are able to derive several performance metrics at the dynamic level. In Fig. 9, we plot, for the different studied scenarios, the evolution of the load in the macrocell and the small cell when the arrival intensity $\lambda_{Tot}$ increases. In addition to the analytical expressions of the load, simulations are also performed so as to validate the former.

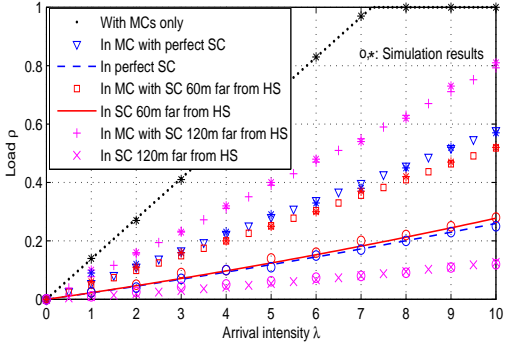

Fig. 9: The load in the small cell and the macrocell.

From Fig. 9, we observe that the load is significantly reduced with the deployment of a small cell even when it is not perfectly located near the hotspot. Moreover, we notice that the load in the small cell is also less than the load in the macrocell. This can be explained by the fact that the traffic hotspot is in the cell edge and interference experienced by users in cell edge is more significant when they are attached to the macrocell compared to users attached to the small cell. Furthermore, we observe that, when the small cell is slightly deviated from its perfect position, the load in the macrocell reduces and the load in the small cell increases. In fact, when the small cell is deployed in the center of the traffic hotspot, it absorbs most of the traffic which leads to more interference on the macrocell users due to the coupling between the macrocell and the small cell.

On the other hand, the load in the perfectly deployed small cell is marginally reduced compared to the load of the small cell slightly deviated from its perfect position (60 meters far from the hotspot). This is due to the increase of the proportion of users having good radio conditions in the perfectly deployed small cell. However, when this latter is slightly far from its perfect position, a significant proportion of traffic becomes with degraded radio condition while its absorption coefficient still remain more or less the same as for perfectly deployed small cell (see Fig. 8). In addition, when the proportion of traffic absorbed by a small cell 120 meters far from the hotspot is significantly reduced (see Fig. 8), its load becomes smaller than the load in perfectly deployed small cell even with users exclusively located at small cell edge. Nevertheless and still in the scenario with small cell 120 meters far from the hotspot, the load of the macrocell is significantly increased compared with the perfect deployment scenario.

In Figs. 10 and 11, we plot the average number of active flows and the mean flow throughput in the macrocell and the small cell. The analytical expressions derived from (28) and (30) are compared with simulation results and we show that the approximations made are accurate for evaluating the system performance. The same conclusions as before regarding the load are recovered for the average number of active flows. First, when the small cell is significantly far from the hotspot, the mean number of active flows in the small cell is reduced comparing to a perfectly deployed small cell. On the other hand, this metric is increased for the macrocell when the small cell is not perfectly positioned near the hotspot since it will take more traffic. Besides, more active flows are observed in macrocell with perfect small cell positioning than in the case of a slightly deviated small cell. In addition to the explanation given in the evaluation of the load, this is related to the large proportion of users that are at the small cell edge and attached to the macrocell (see the absorption coefficient in Fig. 8). And so, these users are highly interfered by the small cell and thus will stay longer in the system to complete their transfers.

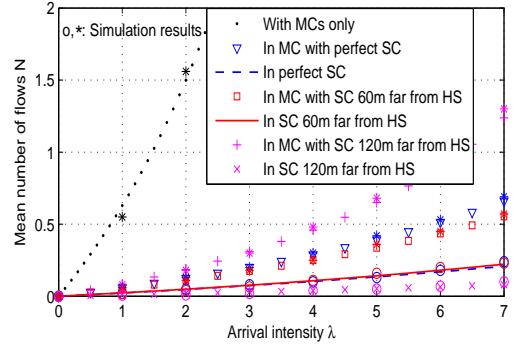

Fig. 10: Average number of active flows.

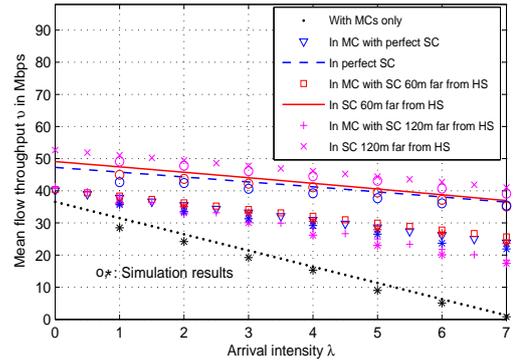

Fig. 11: Mean flow throughput.

Results from Fig. 11 confirm once again the different observations concluded from the load and mean number of flows' analysis. Contrary to the static level results, the mean flow throughput in the small cell and the macrocell is improved when the traffic hotspot is partially absorbed by the small cell (the small cell is not deployed exactly in the center of the hotspot). This can be further explained by the fact that more users will share the small cell resources when it is perfectly deployed and so their transmission rates will be slightly reduced. We also notice that the flow throughput in the small cell deployed far enough from the hotspot is improved conversely to the flow throughput in its macrocell. The increase of the throughput in the small cell is reasonable since the number of active flows is reduced and the cell resources are divided between a smaller number of flows (see the absorption coefficient in Fig. 8).

Finally, we conclude that deploying small cells represents always a good way to offload traffic but improving the efficiency of this solution depends on many constraints in addition to those identified in the static level results. First, the small cell must cover all the hotspot in order to reduce the proportion of macrocell users located near the small cell and experiencing high interference from this later. Second, the capacity of the small cell and the size of its carried traffic are also important to take as an input in the HetNet planning process. Third, in the case where the small cell does not completely cover the hotspot, the interference coming from the small cell on macrocell users represents a crucial limitation leading thus to the need of improving interference mitigation techniques.

## VI. CONCLUSION

We have studied in this paper the impact of deploying a small cell in the presence of traffic hotspot inside a macrocell. We have firstly performed our modeling following a static level analysis to assess the throughput distribution and the absorption coefficient of the small cell. We have secondly investigated the dynamic level analysis where users come to the system at random times and leave it after being served with data rates derived from the static level part.
If one considers analysis at static level only, results show that the efficiency of deploying small cells to offload traffic in the congested macrocell depends mainly on the precision of the traffic hotspot localization process. Moreover, when the hotspot is at the cell edge, the imperfections of the hotspot localization are more tolerated and the system performance can still be improved by deploying small cells as compared to a network composed of only macrocells.
Dynamic level analysis allows us to observe that, for small cell deployments with positive gains, the perfect positioning of the small cell with respect to the hotspot position is sufficient to obtain better performance but it is not sufficient to obtain the best one. This latter depends on several input parameters such as the radio range of the small cell (if it covers all the hotspot or a part of it when this latter is more flat), the small cell capacity and the amount of traffic it carries as well as the level of interference generated by the small cell.

In the future, we will extend this work to include more sophisticated scenarios with the study of moving small cells in the presence of mass events (concert, march, etc.). It concerns the control of small cells' mobility according to traffic hotspot localization method, so as to make this promising technology a cost effective investment. In addition to the requirements imposed to the small cell positioning, other important challenges such as the backhaul capacity must be considered to meet the requirements of ultra dense deployments in 5G networks [19].

## ACKNOWLEDGMENT

The authors would like to thank Y. T. Lin and A. Khlass, of Orange Labs, for their valuable discussions provided in this work.

## APPENDICES

### A- Proof of Proposition 1

In the studied scenario, one has $S = \int_{S^*} dt(r,\theta)$ since no small cells are deployed in the evaluated area. Next, from the throughput CCDF definition given in (7) and based on the monotonicity of $g$, $\mathbb{P}(\eta > l)$ is expressed as follows

$$\mathbb{P}(\eta \geq l) = \frac{1}{S} \int_{S^*} \mathbb{1}\left(g(r) \leq \psi(l)\right) dt(r,\theta)$$
$$= \frac{1}{S} \int_0^R \int_0^{2\pi} \mathbb{1}\left(r \leq g^{-1}(\psi(l))\right) dt(r,\theta)$$
$$= \frac{1}{S} \int_0^\Lambda \int_0^{2\pi} dt(r,\theta)$$

with $\Lambda = \min\left(g^{-1}(\psi(l)), R\right)$ and $\psi(l)$ as defined in (9). Finally, replacing $dt(r,\theta)$ by its expression in (1) and evaluating the integral on $\theta$, we find (10) and complete the proof.

### B- Proof of Proposition 2

First, we replace $\gamma(r,\theta)$ in (7) by its expression given in definition 1. Next, we consider $\Omega$ defined by

$$\Omega = \mathbb{1}\left(g(r) + \kappa r^{2b} |re^{i\theta} - R_s e^{i\theta_s}|^{-2b} \leq \psi(l)\right)$$

If $g(r) \geq \psi(l)$, then $\Omega = 0$. Consequently, since $g$ is an increasing function, $\mathbb{P}(\eta \geq l)$ becomes as follows

$$\mathbb{P}(\eta \geq l) = \frac{1}{S} \int_0^\Lambda \int_0^{2\pi} \mathbb{1}\left(\kappa |re^{i\theta} - R_s e^{i\theta_s}|^{-2b} \leq r^{-2b}\right) \times$$
$$\mathbb{1}\left(g(r) + \kappa r^{2b} |re^{i\theta} - R_s e^{i\theta_s}|^{-2b} \leq \psi(l)\right) dt(r,\theta)$$

with $\Lambda = \min\left(g^{-1}(\psi(l)), R\right)$ and $\psi(l)$ as defined in (9). Subsequently, applying elementary transformations yields to

$$\mathbb{P}(\eta \geq l) = \frac{1}{S} \int_0^\Lambda \int_0^{2\pi} \mathbb{1}\left(\cos(\theta - \theta_s) \leq \min\{g_1(r), g_2(r)\}\right) dt(r,\theta)$$

$g_1(r) < g_2(r)$ if and only if $g(r) < \psi_l - 1$.

Knowing that $g^{-1}$ is an increasing function, one has $g^{-1}(\psi_l) \geq g^{-1}(\psi_l - 1)$. Thus, we obtain (11) which completes the first part of the proof.

The throughput CCDF in the small cell is defined in (8). We replace $\tilde{\gamma}(r,\theta)$ by its expression given in definition 1 and we apply elementary transformations in order to obtain

$$\mathbb{P}(\tilde{\eta} \geq l) = \frac{1}{\tilde{S}} \int_0^R \int_0^{2\pi} \mathbb{1}\left(\cos(\theta - \theta_s) > g_1(r)\right) \times$$
$$\mathbb{1}\left(\cos(\theta - \theta_s) \geq g_3(r)\right) dt(r,\theta)$$

Furthermore, one has $g_1(r) < g_3(r)$ if and only if $g(r) > \psi_l - 1$. Hence, we obtain the expression in (12).

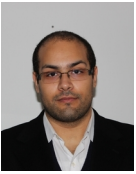

**Aymen JAZIRI** was born in 1989 in Tunisia. He received a double degree from SUP'COM Tunisia and SUPELEC France in 2013 (the Engineering degree in telecommunications and the MSc degree in Wireless communications). He is currently pursuing a Ph.D. degree with Orange Labs jointly with Telecom Sudparis and the University of Paris 6 (UPMC), France. His current research interests include HetNets' planning and performance analysis, traffic modeling in cellular networks, wireless communications and queuing theory.

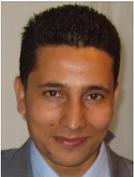

**Ridha NASRI** received his Engineering Diploma and M.S degree from Ecole Superieure des Communications de Tunis, Tunisia, in 2002 and 2004 respectively, and his Ph.D degree in Telecommunications and Informatics from University of Pierre and Marie Curie, France, in 2009. He is currently a senior researcher and technical project manager with Orange Labs, France. From 2009 to 2012, he was with Orange France among the technical steering committee of the Ile de France networks. Before 2009, he served as a senior engineer for network exploitation and planning in different telecommunication companies. He also held a temporary visiting lecturer with Telecom SudParis and Esigetel, France, in 2009 and 2010.

Dr. Nasri has published over 20 refereed journal and conference papers and holds 2 patents. He served in the TPC members of various conferences and a reviewer of different journals (e.g., IEEE Trans. on Communications, IEEE Trans. on Vehicular Technologies, Journal of Integral Transforms and Special Functions). His research interests include Performance analysis of wireless communications, Traffic modeling, Coding and number theory.

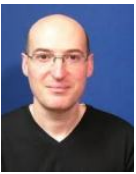

**Tijani CHAHED** has a BSc and MSc in Electrical and Electronics Engineering from Bilkent University, Turkey, a PhD and Habilitation a Diriger des Recherches (HDR) in Computer Science from the University of Versailles and the University of Paris 6, France, respectively. He is currently a Professor in the Telecommunication Networks and Services department, Telecom Sudparis, France. His research interests are in the area of QoS and performance analysis, resource allocation and teletraffic engineering, notably in wireless networks.